\documentclass[showpacs,nofootinbib]{revtex4} 
\usepackage{graphics}
\begin{document}
\title{Tunneling through Color Glass Condensate and~True~Black~Disks}
\author{Alexey V. Popov} 
\email[]{avp@novgorod.net}
\affiliation{Velikiy Novgorod, Russia}
\begin{abstract}
We discover new vacuum solutions of the Jalilian-Marian–Iancu-McLerran-Weigert-Leonidov-Kovner equation, which correspond to center of a gauge group. 
We improve the color glass condensate (CGC) model by an explicit usage of a density matrix. 
Studying scattering of CGC states in an external color field, we observe that an amplitude is naturally expressed via group characters. 
We construct an example that shows how new thin effects may be potentially observed in peripheral collisions. 
We prove that at any parton density a gluonic CGC state does not become a true black disk.
We find a wave function of a true black disk and show that it necessarily contains many quarks.
This result corresponds to the necessity of nonvacuum Reggeon loops in a formation of a true black disk. 
\end{abstract} 
\maketitle 

\newcommand{\ket}[1]{\ensuremath{|#1\rangle}}
\newcommand{\bra}[1]{\ensuremath{\langle#1|}}
\newcommand{\braket}[2]{\ensuremath{\langle#1|#2\rangle}}
\section{Introduction}
Consider high energy processes in the QCD. A hadron at a given rapidity is a composite state of quarks and gluons. 
It is described by a wave function which is some state in the Fock space of partons. 
If we boost a hadron to high rapidity, then its wave function will be changed according to the methodology of evolution equation.
When studying realistic processes, we usually work with an effective coarse-grained theory with a given scale of transverse resolution.
If we change the scale then a wave function will be changed, too. Then we can obtain the DGLAP equation (or usual RG flow for the Lagrangian, if we take a spatially symmetric case).
Usually, the scale of transverse resolution is given by a characteristic momentum transfer $Q$. 
Taking a more coarse scale, we shall calculate a wrong result. Taking a more fine scale, we shall make unnecessary work because in calculations 
of observables we effectively average over modes living far away from the interested momentum transfer.

If at some transverse scale the number of partons per an elementary resolution area is much larger than 1, then we can expect an appearance
of some sort of a statistically effective description. Therefore, the so-called color glass condensate (CGC) model \cite{CGC} was proposed, 
where specific conditions on operator correlators of color charge density was imposed. 
A next natural question is about an explicit partonic structure of the CGC model. 
We want to know what kinds of quantum states produce the CGC description for field correlators.
There is a partonic derivation of the CGC model \cite{Kovchegov_96} 
for a large nucleus, or for an arbitrary dense state \cite{Jeon_04}.
In the work \cite{Jeon_04}, a large number of random color partons in the same place were considered.
In our framework, we associate 
such a place with an elementary cell in the transverse plane. Reference \cite{Jeon_04} concentrates on a distribution of a Casimir operator and 
its main goal is to justify the validity of the classical description, which exploits a classical color charge and a Gaussian distribution as a first approximation.
However, such a description is too peculiar and is devoted only to the construction of an effective theory.
In this paper, we look at the situation from a more general viewpoint where a fast hadron projectile is a quantum system of QCD partons
and a wave function is an element of the Fock space. 
We shall show that Ref. \cite{Jeon_04} implicitly use a density matrix for a wave function of partons.
In this paper, we emphasize the usage of a density matrix.
This allows us to see properties, which lie beyond the scope of classical charges. 
In particular, we show that the CGC states do not tend to a black disk at arbitrary high parton densities.

An important ingredient of high energy QCD is the Jalilian--Marian–-Iancu--McLerran-–Weigert-–Leonidov-–Kovner(JIMWLK) equation \cite{CGC,JIMWLK}.
It describes the evolution of a scattering amplitude of a dilute projectile on a dense target.
The problem of white and black discs was discussed in context of the JIMWLK/KLWMIJ equation \cite{Kovner_06}.
It was shown that there are only two stable points of the evolution: black and white.
In this paper, we show that the situation is more complex.  
On a compact group, instead of a usage of common Fourier transform we must use matrix elements of irreducible representations as a complete basic in a
space of functions. This gives new results which were missed in \cite{Kovner_06}. Particularly, we show that there exist multiple 
stable points of the evolution. In fact, they are the central elements of the gauge group.

In non-Abelian gauge theories, the nontrivial role of the $SU(N)$ center is already known. 
In the thermal QCD the spontaneous broken center symmetry is a property of the deconfinement phase transition
(see \cite{Holland} for a nice review and references therein). 
The so-called Polyakov loop is an order parameter for center symmetry breaking.
The group center, group characters, and the group manifold also emerge in various approaches to the problem of the confinement in QCD.
For an example, in Ref. \cite{Polikarpov} a random walk on the group manifold was observed.

Our new results raise an important question about the structure of a true black disk.
This disk is just a mathematical idealization. The precise definition will be given is Sec. \ref{sec_7}.
In Ref. \cite{Kovner_06}, a black disk was considered as a special choice of 
a weight functional which determines the probability density to find a given
configuration of classical color charges in a projectile. 
The main flaw is the usage of classical charges.
In this paper, we argue that classical charges are not suitable for question about a black disk. 
Our calculation contains only quantum operators which act in the natural projectile Hilbert space.
We show that in order to construct the true black disk we must take many partons in the fundamental representation.

The master plan of our paper is
(1) To explore a partonic structure of the CGC model.
(2) By symmetry reasons, to consider the CGC state as a projectile. 
(3) To study a scattering of the CGC in an external color field.
(4) To investigate the limit of a high parton number. To find points on the group manifold where S-matrix does not vanish. 
(5) To try to find states which in high density limit tend to the true black disk.
(6) To argue that a true black disk must contains many quarks.

This paper is organized as follows.
In Sec. \ref{sec_2}, we review basic properties of the CGC model and observe that there is a natural description of CGC in terms of a density matrix with maximal entropy.   
Then we show that an average of a gauge transformation on the CGC state is equal to the high power of a $SU(N_c)$ group character.
In Sec. \ref{sec_8}, we argue why it is necessary to study a scattering of any state in an external color field.
It will be clear why the method of weight functionals is wrong in a general situation.
In Sec. \ref{sec_3}, we analyze a simple situation of the $SU(2)$ gauge group and find points on a group manifold where the character of the adjoint 
representation has maximal value. In Section \ref{sec_4} we study a more complicated case of the $SU(3)$ gauge group.
In Sec. \ref{sec_9}, we show a scattering process where a center of a gauge group influences a physical observable.
In Sec. \ref{sec_5}, we prove that central points are stable solutions of the JIMWLK equation.
In Sec. \ref{sec_7}, we discuss a question about structure of the true black disk. 
Section \ref{sec_6} contains our conclusions.

\section{Characters in CGC} \label{sec_2}
Consider a hadron state $\ket{\Psi}$ having a large number of partons. Let us take a small area $S_1$ within hadron and
consider a subsystem in $S_1$ as a quantum system. It was motivated in \cite{avp_agk} that in such situations it is useful to use a density matrix for 
the subsystem $S_1$. Moreover, we can expect that in a chaotic environment the entropy tends to a maximally allowed value. 
Maximum of the entropy suggests that all available states are equiprobable and the density matrix is proportional to the unit matrix. 

Consider $N$ partons in color representation $R$. The space of states $H$ of the system is a tensor product of the spaces of state of every parton.
\begin{equation} \label{eq_15}
H=\underbrace{V_R\otimes V_R \otimes \ldots \otimes  V_R}_N=V_R^{\otimes N}
\end{equation}
where $V_R$ is a vector space of representation $R$. A color charge operator is a generator of the $SU(N_c)$ group
\begin{equation} \label{eq_10}
\rho^a=T^a_R\otimes 1 \otimes \ldots \otimes 1 + 1 \otimes T_R^a \otimes 1 \ldots \otimes 1 + \ldots
\end{equation} 
where $T^a_R$ is the generator in the representation $R$. 

Taking the unit matrix in the space $H$ in (\ref{eq_15}), the normalized density matrix of maximal entropy is given by 
\begin{equation} \label{eq_4}
w=\frac{1}{d_R^N}
\end{equation}
where $d_R$ is the dimension of representation $R$. Averages of operators, which act only on $H$, are calculated as
\begin{equation} \label{eq_11}
\langle\hat O\rangle = Sp (w \hat O)=\frac{1}{d_R^N} Sp (\hat O)
\end{equation}
For example, we can calculate the average of the quadratic Casimir operator $C$
\begin{equation}
\langle C \rangle=\langle \rho^a \rho^a \rangle  = \frac{N}{d_R}Sp (T^a_R T^a_R)=NC_R
\end{equation}
where we used the obvious fact that $Sp(T^a_R)=0$. In the subsequent calculation we usually omit representation index $R$.

It should be noted that, in general, we must also consider some 
distribution on $N$. In the statistical limit of large $N$ it usually has Poisson-like behavior. Here we consider a simplified case with
fixed $N$ which is equal to the average parton number in the cell.

Now we shall show how to obtain the usual CGC description in terms of classical color charges with the Gaussian distribution.
We want to calculate averages of arbitrary polynomials of $\rho^a$. 
Using identity $Sp(T^a_R)=0$, formulas (\ref{eq_10}) and (\ref{eq_11}), we directly evaluate 
\begin{equation}
\langle \rho^a \rangle = 0
\end{equation}
\begin{equation} \label{eq_12}
\langle \rho^a \rho^b \rangle = \frac{N}{d_R}Sp (T^a T^b)=\frac{NC_R}{N_c^2-1}\delta^{ab}=\mu^2\delta^{ab}
\end{equation}
\begin{equation} \label{eq_2}
\langle \rho^a \rho^b \rho^c \rangle = \frac{N}{d_R}Sp (T^a T^b T^c)
\end{equation}
\begin{equation}
\begin{array}{rl}
\langle \rho^a \rho^b \rho^c \rho^d \rangle =& \frac{N(N-1)}{d^2_R} \left [  Sp (T^a T^b) Sp(T^c T^d)+ Sp (T^a T^c) Sp(T^b T^d)+ Sp (T^a T^d) Sp(T^b T^c)\right ] \\
& + \frac{N}{d_R} Sp (T^a T^b T^c T^d) 
\end{array}
\end{equation}
In the large $N$ limit we have
\begin{equation} \label{eq_1}
\langle \rho^a \rho^b \rho^c \rho^d \rangle \longrightarrow
\langle \rho^a \rho^b \rangle \langle \rho^c \rho^d \rangle+
\langle \rho^a \rho^c \rangle \langle \rho^b \rho^d \rangle+ 
\langle \rho^a \rho^d \rangle \langle \rho^b \rho^c \rangle 
\end{equation}
and similar formulas for an higher even power of $\rho^a$. In (\ref{eq_1}), we can easily recognize the Wick's rules. Higher
correlators are reduced to all possible pair contractions. 
If we temporarily forget about odd powers, then we can generate averaging rules with the help of the classical Gaussian distribution,
which is described by the following weight functional:
\begin{equation} \label{eq_3}
W[\rho]=\left(\frac{1}{\sqrt{2\pi\mu^2}}\right)^{N_c^2-1}\exp\left(-\frac{1}{2\mu^2} \rho^a \rho^a\right)
\end{equation}
The average classical charge density we define as an average of the absolute value of the following classical variable:
\begin{equation}
\langle |\rho^a| \rangle=\int |\rho^a| W[\rho] d\rho \sim  \sqrt{N}
\end{equation}
Usually, the square root behavior is interpreted as the random walk scenario. The average of the gauge transformation is
\begin{equation} \label{eq_6}
\langle e^{i\alpha_a\rho^a}\rangle=\exp\left(-\frac{1}{2}\mu^2 \alpha^a\alpha^a\right)
\end{equation}
For large $N$ in (\ref{eq_6}), we have more and more narrow peaks near $\alpha_a=0$.

Now consider odd powers of $\rho^a$. It is clear that there is a direct analog of (\ref{eq_1}) for odd powers in the large $N$ limit.
Namely,  at odd $n$ the leading term has only one triple contraction (\ref{eq_2}). 
\begin{equation}
\langle\rho^{a_1}\ldots\rho^{a_n}\rangle=\langle\rho^{a_1}\rho^{a_2}\rho^{a_3}\rangle \langle\rho^{a_4}\rho^{a_5}\rangle \ldots \langle\rho^{a_{n-1}}\rho^{a_n}\rangle + \mbox{permutations}
\end{equation}
Note that $\langle\rho^{a_1}\rho^{a_2}\rho^{a_3}\rangle$ in general is neither symmetric nor antisymmetric. 
Hence, we cannot construct an analog of (\ref{eq_3}) and cannot use classical variables. In order to cure the problem 
the so-called Wess-Zumino term can be used \cite{Kovner_WZ,Hatta}. 
Here we use a different method of generation of averages based on matrix elements of gauge transformations.

Now we switch to the new CGC description which uses functions on the group manifold. 
In Eq. (\ref{eq_10}) the operators $\rho^a$  are generators of gauge transformations. On the other side, the gauge group action on a tensor product is
a tensor product of linear operators which represent group elements
\begin{equation}
U_{A\otimes B}(g)=U_A(g) \otimes U_B(g)
\end{equation}
where $U_A(g)$ is a matrix which represents $g\in SU(N_c)$ in the representation $A$. 
Since any matrix $U(g)$ is a quantum operator in a color space, we can evaluate its average by applying the definition (\ref{eq_11})
with the density matrix (\ref{eq_4}). This gives 
\begin{equation} \label{eq_5}
S[g]=\langle U(g)\otimes U(g) \ldots \otimes U(g)\rangle =\left(\frac{\chi_R(g)}{d_R} \right)^N
\end{equation}
where $\chi_R(g)$ is a character of the representation $R$. Characters can be considered as invariant functions on a group manifold.
Basic properties of characters will be reproduced in Sec. \ref{sec_6}. For $SU(N_c)$, group characters are uniquely defined on diagonal matrixes. 
Any diagonal matrix is defined by its eigenvalues $z_i$. Unitarity leads to $|z_i|=1$. Hence, we have constraint 
\begin{equation}
\left|\frac{\chi_R(g)}{d_R}\right|\leq 1
\end{equation}
So it is clear from (\ref{eq_5}) that 
in the large $N$ limit only the neighborhood of points $|\chi_R|\sim d_R$
have a nonvanishing $S$ matrix.
Moreover, if $\chi_R=d_R e^{i\phi}$ where $\phi\neq 0$, then due to existence of a distribution over $N$, as it was discussed above, such 
points do not survive in the large $N$ limit. So we are primarily interested only in points where $\chi_R=d_R$ and where $S\sim 1$, even in $N\to\infty$ .

Let us introduce canonical coordinates $\alpha_a$ on the group manifold. 
Below we shall simultaneously use two notations for group points: a set of coordinates $\alpha_a$ for an analytical notation, 
and an element of group $g$ for an invariant notation.
Using the coordinates, an average of a gauge transformation has the form
\begin{equation}
 S[\alpha]=\langle e^{i\alpha_a\rho^a} \rangle
\end{equation}
The group multiplication induces two kinds of global vector fields on a group manifold, the so-called left and right
invariant vector fields: $J^a_+$ and $J^a_-$. They act on $S$ as
\begin{equation}
J^a_+ S[\alpha]=i \langle  \rho^a e^{i\alpha_b\rho^b} \rangle
\end{equation}
\begin{equation}
J^a_- S[\alpha]=i \langle e^{i\alpha_b\rho^b} \rho^a  \rangle
\end{equation}
An average of arbitrary power of $\rho^a$ can be calculated by the action of $J^a_\pm$ on function $S[\alpha]$ at the point 
$g=e$ (or $\alpha_a=0$), where $e$ is a identity element of a group.

Physical meaning of $S[\alpha]$ is an elastic $S$ matrix of a scattering of a CGC state in an external color field $\alpha_a$. 
Consider a collision of the CGC state considered as a projectile with some other state named as target. The result (\ref{eq_6}) is valid only near $\alpha_a=0$  because 
(\ref{eq_1}) is valid only if power of $\rho^a$ is much less than $N$. Moreover, (\ref{eq_6}) is not a function on the group manifold. 
Our gauge group is compact and has finite volume. Any function must be at least periodic in the context of canonical coordinates.
The correct value of $S[\alpha]$ for CGC states was calculated in~(\ref{eq_5}). In addition, the result (\ref{eq_5}) is valid for all 
values of $N$. Relatively large $N$ is required only for a search of regions where $S[\alpha]$ is maximal.

Finally, one more note is about the definition of CGC. Originally, it was destined to describe dense states with a large number of partons per 
an elemental cell. However, it is clear that the CGC model is not rigidly restricted to the dense domain.
It is only required that all correlators must vanish, instead of
\begin{equation} \label{eq_13}
\langle\rho^a(x)\rho^b(y)\rangle=\mu^2 \delta(x-y)
\end{equation}
Partons at different transverse cells must be decorrelated. 
Using a density matrix, it is clear that such decorrelation naturally arises in states with large or maximal entropy.
If we select some subsystem, then we can use entanglement entropy. 
A small dilute subsystem, as a part of a large dilute system, may be in a state of maximal entropy due to the combinatoric reason.
There is also a question about the case $x = y$ in (\ref{eq_13}).
We show now that the formula (\ref{eq_13}) may be rewritten in a more natural form, which has no any subtleties like $x=y$.
The right-hand side of (\ref{eq_13}) is entirely related to yields of the normal ordering procedure. Let us recall that
\begin{equation}
\rho^a=T^a_{ij}a^\dag_i a_j
\end{equation}
We wish to know the operator product $\rho^a\rho^b$, where the operators act on same transverse position.
Performing normal ordering, for the fundamental representation we obtain 
\begin{equation}
\rho^a\rho^b = :\rho^a\rho^b:+C_{abc}:\rho^c:+\frac{1}{2N_c}\delta^{ab}\hat N
\end{equation}
where $C_{abc}$ is an unessential group tensor and $\hat N=\sum a^\dag_i a_i$ is the operator of parton number. 
The average $\langle\hat N\rangle$ gives corresponding $\mu^2$ factor in (\ref{eq_13}).
Hence, the CGC model can be reformulated as the following requirement for the state $\ket \Psi$:
\begin{equation} \label{eq_14}
\bra{\Psi} : \rho^{a_1} \ldots \rho^{a_n}: \ket{\Psi}=0
\end{equation} 
for any $n>0$. The formula (\ref{eq_14}) is a natural replacement for (\ref{eq_13}). 

\section{General picture of high energy scattering} \label{sec_8}
In order to motivate the necessity of studying scattering in an external color field  
we propose a general picture of high energy scattering. 
Consider an arbitrary projectile-target scattering. We use Cartesian coordinates with the instant form of dynamic, which allow us to write an explicitly symmetric theory. 
The interaction occurs only in a short time interval near $t=0$.
Let $\ket P$ be a state of the projectile and $\ket T$ be a state of the target.
The initial state at $t=-\infty$ is
\begin{equation}
\ket P \otimes \ket T
\end{equation} 
Note that this state is a pure tensor product. 
We want to know the amplitude to find the system at $t=\infty$ in the given final state $\ket {P'}\otimes \ket{T'}$ which is also a pure tensor product.
The amplitude is a matrix element of the quasielastic $S$ matrix and it is given by 
\begin{equation}
S=\bra {P'} \otimes \bra {T'} \quad \hat S \quad \ket P \otimes \ket T
\end{equation}
where $\hat S$ is the scattering operator.
Any initial state is a quantum superposition 
\begin{equation}
\ket{P}=\sum_k \Psi^P_k \ket{P_k}
\end{equation}
\begin{equation}
\ket{T}=\sum_k \Psi^T_k \ket{T_k}
\end{equation}
where $\ket{P_k}$ and $\ket{T_k}$ are states of pure quasiclassical fields without a quantum interference. 
Also, they play a role of boundary field configurations for path integral at $t=\pm \infty$.
In a theory of point particle, such states are just the set of wave functions $\delta(q-q_0)$.
In a field theory, such states have a fixed field configuration and a wave functional has a form $\delta(\varphi-\varphi_0)$.
In the high energy QCD, as a source of bremsstrahlung field a fast parton with a fixed transverse coordinate generates a classical gauge field without interference.    
Vice versa, this parton propagates in the target field considered as an external field for the parton motion.
If the parton has a smeared wave function, then according to the superposition principle its total amplitude is just 
a linear combination of the quasiclassical amplitudes which is usually expressed via Wilson lines. 
Note that an interference in color indexes can be encoded in terms of classical configurations because
the classical field also has a color index. However, the high energy evolution converts any classical state to a state with an interference,
since new emitted gluons arise with a spreaded wave function.   
A state described by a density matrix also does not generate a pure field.

Following to the usual quasiclassical ideology, we express the action of $\hat S$ on any initial state via a superposition of same action 
on a pure fixed quasiclassical state. In the quasiclassical approximation an amplitude is just  
the exponent $e^{iS_0}$ where $S_0$ is value of a classical action evaluated on a classical solution of motion equations 
which give extremum conditions of the classical action.
So the action of the $S$ matrix has a form
\begin{equation} \label{e1}
\hat S \quad \ket P \otimes \ket T=\sum_{k,s} \Psi^P_k \Psi^T_s \hat S_P(\alpha^T_s) \ket {P_k} \otimes \hat S_T(\alpha^P_k) \ket {T_s}
\end{equation}
The projectile propagates in the target fields and the target propagates in the projectile fields.
Equation (\ref{e1}) is valid since in a high energy scattering the time interval of the interaction is very short and 
a change of the target state does not affect on target fields viewed from the projectile viewpoint. After the collision, causality
disallows an influence of the changed target on the projectile.
The field $\alpha^T_s$ is a classical field generated by the target state $\ket {T_s}$. In the QCD it is a gauge field in the Lorenz gauge $A^\pm_a$ 
emitted from classical parton currents $J^\pm_a$.
Note that the final state in (\ref{e1}) is not a tensor product. It cannot be expressed as $\ket{\tilde P} \otimes \ket{\tilde T}$.

To find an elastic amplitude we set $\ket {P'}=\ket {P}$ and $\ket {T'}=\ket {T}$. It has a form
\begin{equation} \label{e2}
S_{el}=\sum_{k_1,k_2,s_1,s_2} \bar \Psi_{k_2}^P \bar \Psi_{s_2}^T \Psi_{k_1}^P \Psi_{s_1}^T
\bra{P_{k_2}} \hat S_P(\alpha^T_{s_1}) \ket {P_{k_1}} \bra{T_{s_2}} \hat S_T(\alpha^P_{k_1}) \ket {T_{s_2}}
\end{equation}
If we take an asymmetric case with a dense target and a dilute projectile, then $\alpha^P_k$ is small and Eq. (\ref{e2}) converts to
\begin{equation} \label{e3}
S_{el}=\sum_s  \bra{P} \hat S_P(\alpha^T_s) \ket {P}  \left | \Psi_s^T  \right|^2
\end{equation}
This is the well-known representation for an elastic amplitude. The projectile scatters in the external field and the result is 
averaged over various target field configurations with the weight given by target wave function $\Psi_s^T$.
For a general case, Eq. (\ref{e2}) cannot be expressed in the form of Eq. (\ref{e3}). The conception of weight functionals is hardly
inapplicable because the final state is the entanglement state which is not a tensor product. 
Nevertheless, the amplitude $\bra{P_{k_2}} \hat S(\alpha^T) \ket {P_{k_1}}$ is the main building block for the exact expression (\ref{e2}).
This amplitude for fixed projectile states gives a functional over target field.

Concerning the question about a black disk, 
we define a black disk as a state $\ket B$ which for any nonempty projectile $\ket P$ gives exactly zero elastic amplitude.
The usual way is to apply the assumption that $\ket B$ is dense and $\ket P$ is dilute. 
In this case, the black disk condition is converted to a condition on the weight functional $|\Psi_s^B|^2$.
Indeed, the standard CGC model at very large square charge density poses a good approximation to the black disk.
However, the restriction into the dense-dilute case is too crude. The true black disk is defined by the more strong condition.
Namely, $S_{el}$ in Eq. (\ref{e2}) must be equal to zero for any nonempty state $\ket P$. 
Note that this requirement is projectile-target symmetric by the construction.
Currently, we are unable to treat directly the general case. So we restrict ourself to a case where $\ket B$ and $\ket P$ are classical states which were defined previously.
In this case, the black disk condition is 
\begin{equation} \label{e4}
\bra P \hat S_P(\alpha_B) \ket P \quad    \bra B \hat S_B(\alpha_P) \ket B=0
\end{equation}
which is explicitly symmetric, again. 

To satisfy the requirement (\ref{e4}) we cannot claim $\bra P \hat S_P(\alpha_B) \ket P$=0 for any $\ket P$
because such $\alpha_B$ does not exist. Only the second condition can be satisfied  
\begin{equation} \label{e5}
\bra B \hat S_B(\alpha_P) \ket B=0
\end{equation}
for any $\alpha_P\neq 0$. This is exactly why we have studied the scattering of the CGC states in external color fields.
Further, by finding $\ket B$ which satisfies Eq. (\ref{e5}), in Sec. \ref{sec_7} we shall explore the structure of the true black disk. 

\section{$SU(2)$ gauge group} \label{sec_3}
Consider the $SU(2)$ gauge group as a simple example. The diagonal subgroup has one dimension and is generated by the element $T^z$ 
\begin{equation}
g(\varphi)=e^{i\varphi T^z}=
\left( \begin{array}{cc} e^{i\frac{1}{2}\varphi} & 0 \\ 0 & e^{-i\frac{1}{2}\varphi} \end{array} \right)
\end{equation} 
where $\varphi\in[0;4\pi)$. In the adjoint representation we have
\begin{equation}
U_3(g(\varphi))=
\left( \begin{array}{ccc} e^{i \varphi} & 0 & 0\\ 0 & 1 & 0 \\ 0 & 0 & e^{-i\varphi} \end{array}  \right)
\end{equation}
The character is
\begin{equation} \label{eq_17}
\chi_3(\varphi)=1+2\cos \varphi
\end{equation}
Here we see that $\chi_3(2\pi)/3=1$. This point corresponds to $-1\in SU(2)$, which is an element of the center subgroup $Z_2$. Hence, 
in some large nonzero external field corresponded to $-1$ element of group, a CGC state constructed from many 
$SU(2)$ gluons can scatter without interaction at any gluon density. 
And moreover, in Sec. \ref{sec_5}, we shall show that 
the state whose $S[\alpha]$ is localized near the point $g=-1$ is a stable point of the JIMWLK evolution equation.

In general, the irreducible representations of the $SU(2)$ group is labeled by index $l=0,\frac{1}{2},1,\frac{3}{2}\ldots$.
Characters can be directly calculated as
\begin{equation}
\chi_l(\varphi)=\frac{\sin\left(l+\frac{1}{2}\right)\varphi}{\sin \frac{1}{2} \varphi}
\end{equation}
In general, the condition $\chi_l(2\pi)/d_l=1$ holds at any integer $l$.
\section{$SU(3)$ gauge group} \label{sec_4}
In this section, we shall show how to calculate characters of the $SU(3)$ group. 
A first-time reader may skip this section without a loss of the paper completeness. 

The general situation can be considered with the help of the Cartan classification of Lie algebras. A character is an invariant function on a group:
$\chi(g)=\chi(h^{-1}gh)$ for all $g,h\in SU(N_c)$. For any $g$ we can find an element $h$ such that the element $h^{-1}gh$ is diagonal.
All diagonal elements form an Abelian subgroup, Lie algebra which is the equivalent of the Cartan algebra $H$. Often this group is called 
maximal torus. Irreducible representations can be classified by studying vectors in the dual vector space $H^*$ which is a linear space of linear forms on $H$.
For each representation there is a special finite set of forms which are named weight vectors.
For the adjoint representation (the Lie algebra itself) such weight vectors are named roots.
\begin{figure}[h] \centering 
\includegraphics{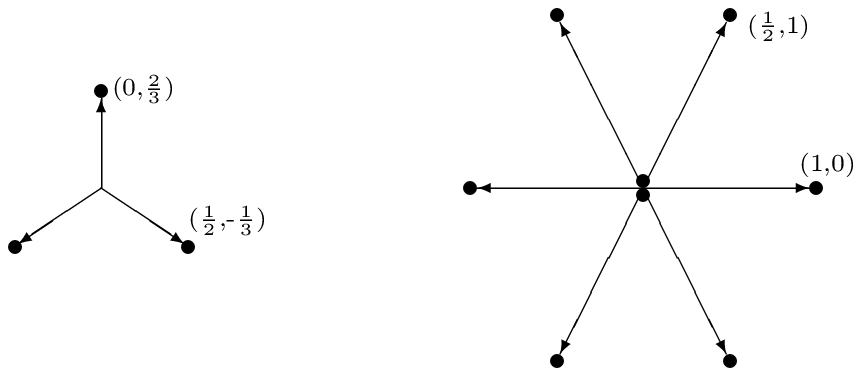} 
\caption{Fundamental and adjoint $SU(3)$ representations.}
\label{fig_1}
\end{figure}
\noindent In Fig. \ref{fig_1}, the fundamental and the adjoint $SU(3)$ representations are shown. Bold points are the weight vectors. 
Roots can reproduce properties of all other irreducible representations. Let $v$ be a vector in some representation and
$T(h)v=\omega(h)v$, where $\omega$ is a weight and $h\in H$. If $\alpha$ is a root, then $T(h)T(E_\alpha)v=T([h;E_\alpha]+E_\alpha h)v=
(\alpha(h)+\omega(h))T(E_\alpha)v$. Hence, the vector $T(E_\alpha)v$ has weight $\alpha+\omega$ or zero.
In addition, we can see that roots set length of all weights. Weights are associated with eigenvalues of $T(h)$. 
So we can directly calculate the character on the diagonal subgroup
\begin{equation}
\chi_R(h)=\sum_k e^{i\omega_k(h)}
\end{equation}
where the implicit exponential map from $H$ to the diagonal group was assumed.
For $SU(3)$ in the fundamental representation we choose the following parametrization:
\begin{equation}
h_1=\frac{1}{2}\left(\begin{array}{ccc} 1&0&0\\0&-1\phantom{-}&0\\0&0&0 \end{array}\right) \qquad \qquad
h_2=\frac{1}{3}\left(\begin{array}{ccc} -1&0&0\\0&-1&0\\0&0&2\end{array}\right)
\end{equation}
The element $h_1$ corresponds to the $x$ axis and $h_2$ corresponds to the $y$ axis in Fig. 1. The diagonal subgroup is generated by
\begin{equation} \label{eq_18}
g(\varphi,\psi)=\exp(i\varphi h_1+i\psi h_2)
\end{equation}
where $\varphi \in [0;4\pi)$ and $\psi \in [0;6\pi)$. The character of the fundamental representation is
\begin{equation}
\chi_3(\varphi,\psi)=e^{i2\psi/3}+2\cos {\frac{\varphi}{2}} e^{-i\psi/3}
\end{equation}
In the adjoint representation, we have eight vectors. Two vectors have zero weight. Six others are generated by symmetries from two
positive vectors: $(1;0)$ and $(\frac{1}{2};1)$. Character of the adjoint representation is
\begin{equation}
\begin{array}{rl}
\chi_8(\varphi,\psi)=&2+e^{i\varphi}+e^{-i\varphi}+e^{i\varphi/2+i\psi}+e^{-i\varphi/2-i\psi}+e^{i\varphi/2-i\psi}+e^{-i\varphi/2+i\psi}=\\
& 2+2\cos\varphi+2\cos\left(\frac{\varphi}{2}+\psi\right)+2\cos\left(\frac{\varphi}{2}-\psi\right)
\end{array}
\end{equation}
In a general case, our calculation is generalized to the well-known Weyl character formula.
Now we want to find points where $\chi_8=8$. There are six pairs $(\varphi,\psi)$, where this condition is satisfied
\begin{equation} \label{eq_7}\begin{array}{ccc}
(0,0) & (0,2\pi)& (0,4\pi) \\ (2\pi,\pi) & (2\pi,3\pi) & (2\pi,5\pi) 
\end{array} \end{equation}
Substituting these points into (\ref{eq_18}), we obtain only three independent diagonal group elements: $1, e^{i2\pi/3}, e^{i4\pi/3}$.
Indeed, they form the center subgroup $Z_3$.
Geometrically, the map from a group to a manifold of linear operators can be viewed as n-fold covering. Points where $\chi_R=d_R$ 
correspond to the kernel of the map. In the $SU(2)$ case, we have the 2-fold covering $SU(2)\to SO(3)$.

\section{An example of tunneling} \label{sec_9}
In order to give a physical manifestation of the explored feature, in this section, we shall construct a gedanken scattering process whose
elastic scattering amplitude is a black disk with ``holes'' corresponding to the central elements of the gauge group.
For the sake of simplicity we consider the $SU(2)$ gauge group. Let the projectile be a small (in the transverse plane) gluonic CGC state
whose average parton density is not required to be high. 
As it was shown in Sec. \ref{sec_8}, to calculate the full elastic scattering amplitude by using the formula (\ref{e3}) we
must assume that the projectile field is small at points where the target is located.
Simultaneously, to see tunneling the target fields at the projectile points must be arbitrary large.
These requirements may be achieved by the following methods: choosing a dense large target, choosing a small dilute projectile,
and increasing the transverse distance between the target and the projectile.
One more simplifying assumption is that the target field $\alpha^T_a(x)$ has a distribution $\left | \Psi_s^T  \right|^2$ 
concentrated strictly near only a one classical field configuration.
In this case, the formula (\ref{e3}) has a simple from
\begin{equation}
S(x)=\bra{P} \hat S_P(\alpha^T) \ket {P}  
\end{equation}
where in the evaluation of the matrix element we can neglect $x$ dependence of $\alpha^T$ if a function $\alpha^T(x)$ is slow varying 
at the projectile points.
Since $\ket P$ is a CGC state, we have the picture discussed in Sec. \ref{sec_2} where a CGC projectile scatters in a constant external color field.
Hence, for $SU(2)$ gluons we can guess 
\begin{equation} \label{eq_16}
S(x)=\left(\frac{1}{3}\chi_3(g(x))\right)^N
\end{equation}
where $N$ is a number of gluon in the projectile, $g(x)$ is the group element corresponded to the group coordinates $\alpha^T_a(x)$.

Strictly speaking, Eq. (\ref{eq_16}) gives a nontrivial contribution only if the projectile is not an overall singlet state. 
Obviously, a singlet state in a constant external field has exactly $S=1$.
Nevertheless, there are no reasons to drop nonsinglet states because a general formalism of scattering must be able 
to treat the whole set of states on same mathematical ground.
In principle, to make the projectile singlet we can add at large $x$ a small patron cluster that discharges the projectile.  
Since at large $x$ the color field is small, new partons have $S=1$ and do not change the projectile $S$ matrix.
The situation can be shown in a more straightforward way. 
Let the projectile be a large gluonic dipole. Its $S$ matrix is
\begin{equation} \label{eq_19}
S(y,x,\alpha)= \frac{1}{3}\chi_3(g^{-1}(y)g(x))
\end{equation}
Let $x$ be fixed. Let us move $y$ to a point where $\alpha\sim 0$ or to the infinity. 
This allows us to approximate  (\ref{eq_19}) as $S\sim \chi_3(g(x))/3$.
So, dealing with a singlet state, we again obtain a expression similar to (\ref{eq_16}). 
Physically, this is a special kind of a border collision where one dipole's parton probes the target and other is a outlying spectator.

\begin{figure}[h] \centering
\includegraphics{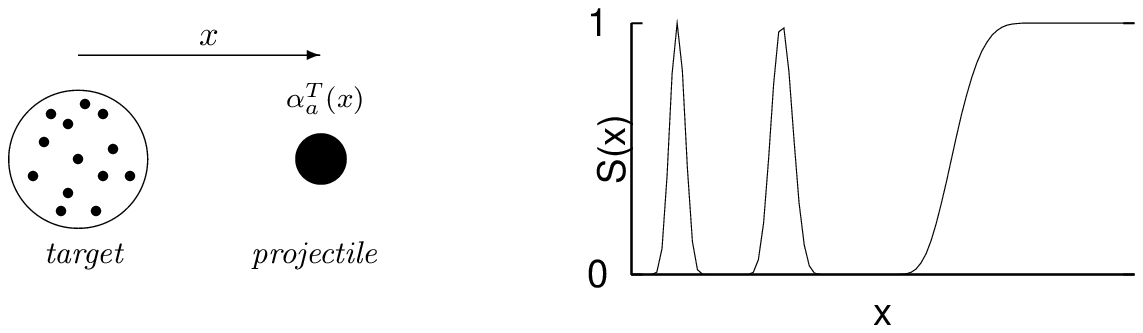} 
\caption{\emph{Left panel:} The scattering setup in the transverse plane. The projectile is a small gluonic CGC state. 
The target is a dense composition of localized patrons. The target field $\alpha^T_a(x)$ is fixed. The impact parameter $x$ is large.
\emph{Right panel:} The schematic plot of $S(x)$. The narrow white peaks arise in the black area.}
\label{graph}
\end{figure}

We discussed in Sec. \ref{sec_8} that a parton with a fixed transverse coordinate and a fixed color state 
generates a classical gauge field without an interference. Two such partons generate a pure field, too.
So we construct the target as a dense color multipole. Namely, the target is an aggregation of a large number of independent partons which
generate large, undistributed total field at the projectile. We collect our scattering setup in the left part of Fig. \ref{graph}.

To calculate the scattering amplitude (\ref{eq_16}) we need to know the target field $\alpha^T_a(x)$. 
We estimate it as a color multipole which at large distances has a power-like asymptotic. 
Taking a fixed transverse direction of the vector $\vec x$, we guess 
\begin{equation}
\alpha_a^T(x)=\frac{c_a}{x^\lambda}
\end{equation}
where $x=|\vec x|$, and $c_a$ is a color factor which describes the color charges in the target. 
By unitary transformation in $SU(2)$, the matrix $c_a T^a$ can be diagonalized as $c \sigma^3/2$.
Hence we can apply Eq. (\ref{eq_17}) which gives
\begin{equation}
S(x)=\left( \frac{1+2\cos (c/x^\lambda)}{3} \right)^N
\end{equation}
A schematic plot of $S(x)$ has been drawn on the right part of Fig. \ref{graph}. 
At large $x$ there is a white area $S=1$. However, at low $x$ (black area) there are narrow white peaks which are a new result. 
The original CGC model has no such peaks. 
The origin of these peaks is the central elements of the gauge group. 
When the field $\alpha^T$ is near a central element, the scattering amplitude fast rises.
A physical consequence of this result is that at some exceptional impact parameters the inelastic processes is strongly suppressed, 
since there is only elastic channel at $S=1$. This explains the exploited term ``tunneling''. 
The phenomenon occurs only for a non-Abelian gauge group. For an Abelian case, characters always obey $|\chi|=1$.

\section{New vacuum solutions of JIMWLK equation} \label{sec_5}
The JIMWLK evolution equation \cite{JIMWLK} can be expressed in terms of invariant vector fields on a group manifold \cite{Weigert_05}. 
We also use the notation taken from Ref. \cite{Kovner_05}. The JIMWLK equation is
\begin{equation} \label{eq_8}
\frac{dS[\alpha]}{dY}=\int\limits_{zxy}^{\phantom{z}} K_{xyz}
\left( -J_+^a(x)J_+^a(y) -J_-^a(x)J_-^a(y)+2V_{ba}(z)J_+^b(x)J_-^a(y)  \right) S[\alpha]
\end{equation}
\begin{equation}
K_{xyz}=\frac{g^2}{(2\pi)^3}\frac{(\vec z-\vec y)(\vec z-\vec x)}{(\vec z-\vec y)^2(\vec z-\vec x)^2}
\end{equation}
where $V_{ba}(z)$ is a gluon scattering amplitude in an external field $\alpha_a$.
We have shown in the previous section that at the points $g_0:$ $\chi_R(g_0)=d_R$ the operator $U_R(g)=1$. 
Consider a model functional 
\begin{equation}
S[\alpha]=\prod_x\delta(g(x)-g_0)
\end{equation}
where the expression $g-g_0$ is formal and has sense only in coordinates.
If $S[\alpha]$ is localized near $g_0$ in the whole
transverse plain, then in Eq. (\ref{eq_8}) we have $V_{ba}(g)\to V_{ba}(g_0)=\delta_{ba}$.
The left and right invariant vector fields are related to each other by the adjoint representation
\begin{equation}
J_+^a f(g)= \left. \frac{\partial}{\partial h^a}f(hg)\right|_{h=0}= 
\left. \frac{\partial}{\partial h^a}f(gg^{-1}hg)\right|_{h=0}=V_{ab}(g)J_-^b f(g)
\end{equation}
Using the fact $V_{ba}(g_0)=\delta_{ba}$ once more, we have
\begin{equation} \label{eq_9}
J_+^a \delta(g-g_0)=J_-^a \delta(g-g_0)
\end{equation}
where instead of $\delta$ function we can take any function localized near $g_0$. 
Substituting property (\ref{eq_9}) into (\ref{eq_8}), we see that $\delta(g-g_0)$ is a solution of the JIMWLK equation
with zero eigenvalue. Strictly speaking, such solution is not physical because a physical state must obey condition $S[0]=1$.
Though, physical states may have a nonzero projection on $\delta(g-g_0)$. 
For example, in Secs. \ref{sec_3} and \ref{sec_4}, we have seen that the gluonic CGC state indeed has the nonzero projection.    
In relation with Ref. \cite{Kovner_06}, where an existence of only two physical vacuum states: black and white was motivated, 
our new states are grey due to a possibility of tunneling through some external nonzero field. 

\section{True black disks} \label{sec_7}
In the previous sections, we have shown that the CGC model does not contain a black disk, since in the special external fields 
any CGC state has a nonzero $S$ matrix. Next, we shall step beyond the CGC model. 
It is natural to ask what projectile states have an exactly zero $S$ matrix on a whole group manifold instead of a unit element. 
Such state is a natural candidate to a black disk.

Usually, a black disk is defined as some dense state which absorbs any projectile. In this section, we consider a reverse case where a black disk
is a fast projectile and scatters in an external color field.  
Any hardon in the QCD can be described by a wave function represented as certain vector in the Hilbert space.
Consider a collision of a black disk state $\ket{\Psi}$ with a some target. We take 
a small black disk and a very large and dense target. 
We suppose that the target color field is much stronger than the projectile field. This does not suppose that the target is a black disk.
The key issue here is that, freely choosing target, we can construct an external color field as we want.
In the considered case, the projectile elastic $S$ matrix has the form
\begin{equation}
S=\int \bra{\Psi} e^{i\int \alpha_a(x)\rho^a(x) dx} \ket{\Psi} W[\alpha] D\alpha
\end{equation}
where $\alpha_a(x)$ is a classical field generated by some quasiclassical element of the target wave function, $\rho^a(x)$ is the operator of color charge 
(the generator of gauge transformations) which acts on the projectile state, 
$x$ is a transverse position. The functional $W[\alpha]$ is a weight functional which is obtained from the target wave function.
Since the target is arbitrary, we shall not perform the average on $\alpha_a$ and consider a configuration with fixed $\alpha_a$.
So the elastic projectile $S$ matrix equals to the functional S[$\alpha]$.

In the previous speculations, there was one mistake. Though the field $\alpha_a(x)$ can be very large, due to compactness of the gauge
group the scattering amplitude is bounded. So, in general, we have no way to compare fields $\alpha_a(x)$ in the context of scattering.
Nevertheless, we define a black disk state as a state which has a zero $S$ matrix if $\alpha_a(x)\neq 0$ in the transverse area where the black disk is located, 
and which has $S=1$ if $\alpha_a(x)=0$ in the same area.  
We shall name this state as true (or extremal) black disk because 
usually in current literature more weak definitions are used.
We define
\begin{equation} \label{f_2}
S[\alpha]=\bra{\Psi} e^{i\int \alpha_a(x)\rho^a(x) dx} \ket{\Psi}
\end{equation}
where it should be stated that the $x$ integral in (\ref{f_2}) is bounded by points where 
$\rho^a(x)\ket{\Psi}\neq 0$.
Since the true black disk must be black for any target and any $W[\alpha]$, we must claim
\begin{equation} \label{eqb_1}
S[\alpha]=Z\prod_{x,a} \delta(\alpha_a(x))
\end{equation}
in the sense of functional generalization of the $\delta$ function. 
The factor $Z$ is the normalization constant which is not important for us.
It was proposed in \cite{Kovner_06} to use variables $\omega^a(x)$ which
arise from Fourier transform of $S[\alpha]$ 
\begin{equation}
S[\omega]=\int S[\alpha] e^{i\int \alpha_a(x) \omega^a(x) dx} D\alpha
\end{equation}
For the infinite black disk this gives a constant functional. The variables $\omega^a(x)$ are usually called classical color charge.
But this method is applicable only for small $\alpha_a$. For arbitrary $\alpha_a$ we should not forget about $SU(N_c)$ compactness. 
Via the exponential map $e^{i\alpha_a \rho^a}$ (a map from a Lie algebra to a representation) the variables $\alpha_a(x)$ can be viewed as canonical 
coordinates on a group manifold. The integration over $\alpha_a(x)$ must be considered as the integration over a group manifold over Haar
measure\footnote{For the diagonal Killing form $Sp(T^aT^b)\sim\delta^{ab}$ and small $\alpha_a$ the Haar measure degenerates to the usual form $\prod_a d\alpha_a$.}. 
Instead of $\alpha_a(x)$ we introduce the variables $g(x)$ which is a map from the transverse plane to the group manifold $g(x): R^2\rightarrow G$.
The claim (\ref{eqb_1}) can be rewritten as
\begin{equation}
S[g]=\prod_x \delta(g(x)-e)
\end{equation}
where $e$ is the identity element of the group $G$. The function $\delta(g-e)$ can be viewed as a function on the group manifold with localized supremum.
The expression $g-e$ is rather formal and has a sense only in the coordinates. 
In general, we define $\delta$ function as a linear functional over functions on a group manifold: 
$\int \delta(g-e) f(g) dg = f(e)$

On a compact group instead of using the Fourier transform we must use the matrix elements of the irreducible representations as a complete basic for
space of functions. The well-known Peter-Weyl theorem says that function
\begin{equation}
b^R_{ij}(g)=\sqrt{d_R} \pi^R_{ij}(g)
\end{equation}
forms an orthonormal and complete basic in the Hilbert space $L^2(G,dg)$:
\begin{equation}
\int \overline{b^Q_{kl}(g)} b^R_{ij}(g) dg=\delta_{RQ} \delta_{ki} \delta_{lj}
\end{equation}
The letters $R$,$Q$ denote irreducible representations; $d_R$ is dimension of a representation; and $\pi^R_{ij}(g)$ is a matrix 
element of the representation $R$ on the group element $g$.

A function $f(g)$ is called central function (or class function) if $f(hgh^{-1})=f(g)$ for any $h\in G$. It is easy to show that the 
function $\delta(g-e)$ is central due to invariance of the Haar measure
\begin{equation}
\int \delta(hgh^{-1}-e) f(g) dg=\int \delta(g-e) f(h^{-1}gh) dg=f(e)
\end{equation}
A direct consequence of the Peter-Weyl theorem is that characters of irreducible representations form an orthonormal and complete basic 
in the Hilbert space of central functions. Characters are given by
\begin{equation} \label{eqb_4}
\chi_R(g)=\sum_i \pi^R_{ii}(g)
\end{equation}
\begin{equation}
\int \overline{\chi_R(g)} \chi_Q(g) dg=\delta_{RQ}
\end{equation}
Any central function $f(g)$ can be decomposed in series on characters 
\begin{equation}
f(g)=\sum_R f_R\chi_R(g)
\end{equation}
\begin{equation}
f_R=\int \overline{\chi_R(g)}f(g) dg
\end{equation}
Here we are interested in a decomposition of the $\delta$ function.
\begin{equation} \label{eqb_2}
\delta(g-e)=\sum_R d_R \chi_R(g)
\end{equation}
where we used the obvious fact that $\chi_R(e)=d_R$. It should be emphasized that in a case of the gauge group $SU(N_c)$ at $N_c>2$ there are
representations which are not equivalent to its complex conjugate. For example, a quark is not an equivalent representation for an antiquark.  
Such representations have complex characters, but they are not purely imaginary near $e$. The formula (\ref{eqb_2}) is a rather 
mathematical idealization. Infinite series are irrelevant for the physical reasons. In the real world, the $\delta$ function must be smeared
near $e$. So the black disk $S$ matrix is not zero for very small fields $\alpha_a$. This is the equivalent of a truncation of the series (\ref{eqb_2})
which, at first sight, is  hardly divergent due to the $d_R$ growing. The convergence of (\ref{eqb_2}) is achieved by alternating characters
values at nonsingular points.
High representations in (\ref{eqb_2}) correspond to high frequencies in the usual Fourier analysis. This happens due to the growing of
the second Casimir operator $D_2$ when dimension of a representation grows\cite{Jeon_04}. The second Casimir operator on functions is 
the Laplace operator on the group manifold. The eigenvalues of the Laplace operator correspond to square of frequencies vector in a flat case. 
It well known from the wave packet calculations that this high frequencies suppression 
is equivalent to spreading of a wave packet. In order to obtain a more narrow packet we must include higher harmonics.
As an example, in Sec. \ref{sec_3}, the characters of the group $SU(2)$ were calculated explicitly.

Characters have useful properties. 
Cyclicity of the trace in (\ref{eqb_4}) gives $\chi(hg)=\chi(gh)$.
The group multiplication induces two kinds of global vector fields, which are called
left and right invariant vector fields: $J_+^a$ and $J_-^a$. Namely, there is a linear map from Lie algebra to vector fields on the group manifold.
Acting on characters, the invariant vector fields give equal results.
\begin{equation} \label{eqb_3}
J_+^a\chi(g)= \left. \frac{\partial}{\partial h^a}\chi(hg)\right|_{h=0}=
\left. \frac{\partial}{\partial h^a}\chi(gh)\right|_{h=0}=J_-^a\chi(g)
\end{equation}
From unitarity of representations we have
\begin{equation}
\left.J^a_\pm \chi(g) \right|_{g=e}=0
\end{equation} 
In Sec. \ref{sec_5}, we have shown that the functional (\ref{eqb_1}) obeys the JIMWLK equation.
This means $dS[\alpha]/dY=0$. Strictly speaking, the JIMWLK equation is valid only in the dilute regime, so
our calculation has only demonstrative purposes. 

The JIMWLK Hamiltonian is a linear operator. If we take its eigenvalue $-\lambda$,  then at high $Y$ the corresponding 
eigenfunctional disappears as $\exp(-\lambda Y)$. The true black disk has zero eigenvalue. So it is a stable state of the evolution.
This is not true at the black disk border where there is a transition region (crossover) to the white state. Obviously, any
black disk must grow when we go to high $Y$. Hence, at the black disk border an interesting process will occur. Emitted quarks and gluons 
will saturate the border by higher characters, thus the border will become black.

The next question is about a possible partonic interpretation of the characters $\chi_R$. The most obvious method is to use a density matrix, again. 
Consider one parton of color charge $R$. Let the density matrix of the parton color space be proportional to 1 (maximum entropy state).
The normalization factor is simple $d_R^{-1}$. Then the elastic scattering amplitude in an external field $\alpha_a$ is 
\begin{equation}
\langle e^{i\alpha_a T^a_R}\rangle=\frac{1}{d_R}Sp\left(e^{i\alpha_a T^a_R}\right)=\frac{1}{d_R}\chi_R(g)
\end{equation}
where the averaging is assumed over the density matrix.
The average color charge within the black disk is zero:
\begin{equation}
\langle\rho^a\rangle=\left.J^a \delta(g-e) \right|_{g=e} =0
\end{equation}
This behavior corresponds to the behavior of CGC states, where, in addition, it is required that $\langle\rho^a(x)\rho^b(y)\rangle=0$ for $x\neq y$.
In difference from the CGC model, in the our black disk state the averaged charge square  $\rho^a(x) \rho^a(x)$ is divergent.

It follows from (\ref{eqb_2}) that it is absolutely necessary to have many quarks and antiquarks in the black disk wave function. Many representations cannot be
obtained via a reduction of tensor products of the gluon representation. For example, in the $SU(3)$ case such representations are 
$\textbf 3,\bar\textbf 3,\textbf 6 \ldots$ Currently, we are not able to predict how such a scenario actually can be realized in nature,
but the necessity of existence of partons in the fundamental representation in the \textbf{true} black disk state was proven. 
So we can suggest the following hypothetical scenario. Starting from a dilute projectile such as a dipole, when boosting it to high $Y$ we predominantly 
generate many new gluons in the wave function. When the gluon density becomes high, then there will be a large 
probability of emitting soft quarks into the new opened phase space. This corresponds to the mechanism of nonvacuum Reggeon in QCD.
See \cite{Kovchegov_03} for a modern example of a quark propagation to a low rapidity.
Usually, such processes are kinematically suppressed in comparison with a gluon emission, but we can expect that in a case of a high number
of sources it can be very probable. A fast gluon converts into a fast quark and a slow antiquark and vice versa. 

Another possible interpretation comes from the running coupling effect studying \cite{Kovchegov_06}. At NLO there are diagrams 
with quark bubbles. Usually, such diagrams are associated with change in the effective QCD coupling constant. 
However, in the context of the current paper, we see that such diagrams can really saturate the black disk wave function by the required fundamental
representations. So the considered true black disk really may be an endpoint of an evolution equation constructed with higher order contributions of the perturbation theory.

The fact that in order to obtain appropriate characters we took the density matrix of maximum entropy naturally corresponds to 
an expected chaotic behavior of a wave function boosted to high energy. 

\section{Conclusion and discussion} \label{sec_6}
In this paper, we have found the new vacuum solutions of the JIMWLK equation and have explored the structure of the true black disk.
Although it is hard to find any relation of this result
to available experimental observables, it gives a new theoretical insight to the QCD asymptotic at very high energies and densities.
In perspective, we can use geometry on a group manifold as a powerful mathematical tool, which helps to improve our understanding of high
energy QCD.
In particular, we have shown that the CGC states do not tend to a black disk. 
We have shown that the true black disk has a more complicated structure and it must be constructed from the whole set 
of irreducible representations which may come from a composite state of initial partons.

One can be confused by the fact of existence of a large number of quarks in the black disk wave function. 
However, we stress here that this notion is about an ideal black disk, which is defined by the rigorous mathematical definition (\ref{eqb_1}).
Such black disks may arise only at infinity high $Y$ as a final state of the evolution. Far before infinity $Y$ we shall deal 
with some approximations, which are experimentally indistinctive from the true black disk.

Let us discuss a relation of our method to a saturation scale.
In a description of an experimental data a so-called saturation scale $Q_s(Y)$ is widely used.
It is a useful quantity that delimits the high and low density regions of high energy QCD. 
Experimentally, it is related to a dipole scattering amplitude.
A dipole size, at which an amplitude becomes saturated, has the scale $1/Q_s$. 
Before the saturation there is a geometrical scaling where some observable depends only on the dimensionless quantity $Q^2/Q^2_s(Y)$.
In the original CGC model $Q_s$ is associated with the average square charge density $\mu^2$, which enters in (\ref{eq_12}),(\ref{eq_6}).
Because of the evolution, the parton density grows and $Q_s$ grows, too.
In the context of the current paper we used the clear partonic interpretation of a saturation scale. 
Namely, this is a scale, at which the average parton density is near 1. 
Quantitatively, it is proportional to square root from the transverse parton density, which is taken at some initial scale $Q_0$
(which is needed to reflect the DGLAP evolution).
If we go to high $Y$, then the parton density will grow and $Q_s$ will grow, too.
Making our conclusion, we see that $Q_s$ is too crude a characteristic of processes.
To see more thin effects we have to use more susceptible quantities. 

\section*{Acknowledgments}
I thank N.V. Prikhod'ko for feedback and useful remarks. I am also grateful to A.V.~Leonidov for interesting discussions.

\end{document}